\documentclass[12pt]{article}
\usepackage[dvips]{graphicx}
\usepackage{epsfig}

\renewcommand{\bar}[1]{\overline{#1}}
\newcommand{\VEV}[1]{\left\langle{#1}\right\rangle}
\newcommand{\ket}[1]{\vert\,{#1}\rangle}

\begin{document}
\begin{flushright}
{\small
SLAC--PUB--11611\\
UK/TP-05-16 \\ 
January 2006 \\}
\end{flushright}

\begin{center}
{\Large
\bf Discrete Symmetries on the Light Front and \\[0.5ex]  a General Relation
Connecting Nucleon Electric \\[1ex] Dipole and Anomalous Magnetic Moments
}\footnote{Work supported in part by the U.S. Department of Energy under 
contracts DE--AC02--76SF00515 and DE--FG02--96ER40989
and the KOSEF (Korea Science and Engineering Foundation).} \vspace{1.5cm}
%DE-FG02-96ER40989 and DE-FG01-00ER45832.

\bigskip

Stanley J. Brodsky\\
Stanford Linear Accelerator Center, Stanford University, Stanford,
California 94309

\bigskip

Susan Gardner\\
Department of Physics and Astronomy, University of Kentucky,
Lexington, Kentucky 40506-0055

\bigskip

Dae Sung Hwang\\
Department of Physics, Sejong University, Seoul 143--747, Korea

\bigskip
%\date{\today}
%{\bf January 4, 2006} 
\bigskip
\end{center}
\bigskip

{\noindent 
We consider the electric dipole form factor, $F_3(q^2)$, as well as the
Dirac and Pauli form factors, $F_1(q^2)$ and $F_2(q^2)$, of the
nucleon in the light-front formalism. 
We derive an exact formula for $F_3(q^2)$ to complement those known for
$F_1(q^2)$ and $F_2(q^2)$. 
We derive the light-front representation of the discrete
symmetry transformations 
and show that time-reversal- and parity-odd effects are captured 
by phases in the light-front wave functions. We thus determine that 
the contributions to $F_2(q^2)$ and 
$F_3(q^2)$, Fock state by Fock state, are related,
independent of the fundamental mechanism through
which CP violation is generated.
Our relation is not specific to the nucleon, but, rather, is
true of spin-$1/2$ systems in general, be they lepton or baryon.
The empirical values of the anomalous magnetic
moments, in concert with empirical bounds on the associated
electric dipole moments, can better constrain theories of CP violation.
In particular, we find that 
the neutron and proton electric dipole moments echo the  
isospin structure 
of the anomalous magnetic moments, $\kappa^n \sim - \kappa^p$.  
}

\vfill\vfill
\newpage

\section{Introduction} 

The electric dipole moments of particles such as the neutron, electron, 
muon, or neutrino, 
provide important windows into the 
fundamental origin of CP violation at the Lagrangian level. 
The underlying source, or sources, of CP violation in Nature 
could arise in any
of a number of ways.  Such sources include not only the phase structure of the
Cabibbo-Kobayashi-Maskawa (CKM) 
matrix~\cite{CKM}, which describes
quark mixing and provides CP violation in the Standard Model, 
but also the phase structure of the lepton-mixing matrix~\cite{MNS}, 
as well as 
flavor-diagonal, CP-violating
interactions, as could occur in theories with extended Higgs sectors, such
as in supersymmetry~\cite{susy}. 
The fundamental theory then leads to effective, higher-dimension, 
CP-violating operators, such as the antisymmetric 
product of three gluonic~\cite{weinberg} or $SU(2)_L$  field 
strengths~\cite{Barr:1990vd} 
or the electric-dipole interaction
$\overline \psi \gamma_5 \sigma_{\mu \nu} F^{\mu \nu} \psi$. 
Thus far experiment has provided upper bounds on the magnitude 
of the electron~\cite{edmlimite} and neutron~\cite{edmlimitn} 
electric dipole moments; current limits 
imply 
that models with weak-scale supersymmetry and 
${\cal O}(1)$ CP-violating parameters can 
produce
electric dipole moments 
significantly in excess of experimental bounds~\cite{edmrev05}. 
New, improved experiments, as in Refs.~\cite{demille,nedmfuture,Raedmfuture}, 
have the capacity to sharpen 
such constraints severely; it is our purpose 
to consider the ramifications of such improvements for theories
of CP violation. 

An essential question is how to relate 
the electric dipole moments of leptons and  
baryons to the CP-violating parameters of the 
underlying theory. 
The light-front Fock expansion~\cite{dirac1949,IMF} 
provides an exact Lorentz-invariant representation
of the matrix elements of the electromagnetic current in terms of the 
overlap of light-front wave functions~\cite{DrellYan,BrodskyDrell,West} --- 
note Ref.~\cite{BPPrev} for a comprehensive review. 
The current takes an elementary form in the light-front %(LF) 
formalism 
because, in the interaction picture, the full Heisenberg current can
be replaced by the free quark current $J^\mu(0)$, evaluated at the 
light-cone time $x^+=0$. 
As first shown by Drell and Yan~\cite{DrellYan,BrodskyDrell}, 
such matrix elements are most readily evaluated 
from the matrix elements of the current $J^+(0)$ 
in the $q^+=0$ frame. 
In contrast to the covariant Bethe-Salpeter formalism, 
familiar from the analysis of hydrogenic bound states in 
quantum electrodynamics (QED)~\cite{bethe}, 
in the light-front formalism 
one does not need to sum over the contributions to the current from an 
infinite number of irreducible kernels.  
Indeed, the evaluation of the current matrix elements  
is intractable in the standard, i.e., instant-form, Hamiltonian formalism, 
since the wave functions are frame-dependent, 
and as one must also take into account 
all interactions of the current with vacuum fluctuations~\cite{BPPrev}.

The light-front formalism is thus ideally suited for computing 
electromagnetic properties of both elementary and composite 
states. The electric dipole form factor is rendered nonzero by 
time-reversal-odd and parity-odd effects 
in the light-front Fock-state wave functions  
themselves. This could occur, for example, at a fundamental level through
higher Fock states which explicitly contain three generations of quarks.
Alternatively, one can integrate out 
the effects of the heavy particles to obtain 
an effective chiral theory in which 
the light-front wave functions are expressed 
in terms of effective meson and baryon degrees of freedom.

In this paper we evaluate the electric dipole form factor, $F_3(q^2)$, 
in the light-front formalism and compare it with the well-known expressions
for the Dirac and Pauli form factors, 
$F_1(q^2)$ and $F_2(q^2)$~\cite{BrodskyDrell}. 
In order to explore 
the structure of the resulting expression for $F_3(q^2)$ 
we explicitly construct and classify the action of 
the discrete operators corresponding to the time-reversal, parity, 
and charge-conjugation transformations acting on wave functions 
realized from quantization on the light-front.  
We then construct the general form of a light-front wave function 
in the presence of fundamental CP violation. 
This, in turn, leads to a model-independent 
relation which connects the time-reversal-odd and parity-odd 
$F_3(q^2)$ form factor to the Pauli form factor 
$F_2(q^2)$, Fock state by Fock state --- for any spin-{1/2} system. 
Thus we are able to relate 
the contribution of a particular Fock state to the electric 
dipole moment to a corresponding contribution to the anomalous 
magnetic moment.   
At $q^2=0$, the universal relation for a spin-$1/2$ baryon is 
\begin{equation}
d_i = 2 \kappa_i \tan \beta_i \,,
\label{unirel}
\end{equation}
where  repeated indices are not summed and $i$ denotes the contribution
of Fock state $i$. 
Note that the electric dipole moment $d$ is 
$d\equiv\Sigma_i d_i = (e/M)F_3(0)$
 and that the anomalous magnetic moment $\kappa$ is  
$\kappa\equiv \Sigma_i\kappa_i = (e/2M)F_2(0)$, 
where $e=|e|$ is the fundamental
unit of electric charge and $M$ is the proton mass. 
The parameter $\beta_i$ is the CP-violating phase appearing in 
Fock state $i$ of the light-front 
wave function for the baryon of interest. Although it has long
been recognized that 
the hadronic matrix element yielding the neutron electric
dipole moment must be commensurate in size, up to 
CP-violating effects,  
to that of the anomalous magnetic
moment~\cite{bigiuralt}, ours is the first construction of
a general equality based on first principles. 
A relationship of this kind has also been noted by 
Feng, Matchev, and Shadmi in their study of the electric-dipole
and anomalous-magnetic moments of the muon in 
supersymmetric models~\cite{Feng:2001sq}, though 
we find our Eq.~(\ref{unirel}) to be of more general validity. 
Indeed, the connection is general and holds for any spin-{1/2} state, be it 
charged lepton, 
neutrino\footnote{The connection is nontrivial only in the case
of a Dirac neutrino.}, quark, or baryon, irrespective of
the sources of CP violation. 
We proceed to examine its implications 
for constraints on models of CP violation 
before concluding with a summary and outlook.

\section{The Light-Front Fock Representation}
The light-front Fock expansion of any hadronic 
system is constructed by quantizing 
quantum chromodynamics (QCD) at fixed light-cone time 
$x^+ = x^0 + x^3\,\,$\footnote{We summarize our conventions
in the Appendix.}, with $c=\hbar=1$, and forming the
invariant light-cone Hamiltonian $H_{LC}$: $ H_{LC} = P^+ P^- - {\mathbf
P}_\perp^2$~\cite{dirac1949,IMF,BPPrev}.  The momentum generators 
$P^+$ and $\mathbf{P}_\perp$ are kinematical, 
so that they are 
independent of interactions~\cite{dirac1949}. 
The generator $P_+ = P^-/2 = i {
\partial/\partial}x^+ $ 
gives rise to light-cone time translations. 
In principle, solving the 
$H_{LC}$ eigenvalue problem gives the entire mass spectrum of the
color-singlet hadron states in QCD, together with their respective
light-front wave functions.  In particular, the proton state satisfies
$H_{LC} \ket{\psi_p} = M^2 \ket{\psi_p}$, 
where $\ket{\psi_p}$ is an expansion in multiparticle Fock states. 
The resulting equations can be 
solved, in principle, using the discretized light-cone quantization
(DLCQ) method~\cite{Pauli:1985ps}.   
A recent example of nonperturbative light-front solutions 
for a $3+1$ theory is given in Ref.~\cite{Brodsky:2005yu}.  
The connection to the Bethe-Salpeter formalism 
is described in Ref.~\cite{Lepage:1980fj}, and explicit examples
thereof are given in Ref.~\cite{Brodsky:2003pw}. 
In the case of elementary fields such as the electron, 
one can construct the Fock space in perturbation theory.

The expansion of 
the proton eigenstate $\ket{\psi_p}$ in QCD 
on the eigenstates, $\{\ket{n} \}$, of the
free light-cone 
Hamiltonian gives the light-front Fock expansion:
\begin{eqnarray}
\left\vert \psi_p (P^+, {\mathbf{P}_\perp}, S_z)\right> &=& 
\sum_{n, \lambda_i\in n}\
\int \prod_{i=1}^{n}
\left(  {{\rm d}x_i\, {\rm d}^2 {\mathbf{k}_{\perp i}}
\over  2 \sqrt{x_i}\, (2 \pi)^3}\ \right) \,
16\pi^3 \delta\left(1-\sum_{i=1}^{n} x_i\right)\,
\delta^{(2)}\left(\sum_{i=1}^{n} {\mathbf{k}_{\perp i}}\right)
\nonumber
\\
&& \qquad \rule{0pt}{4.5ex}
\times \psi^{S_z}_{n/p}(x_i,{\mathbf{k}_{\perp i}},
\lambda_i) \left\vert n;\,
x_i P^+, x_i {\mathbf{P}_\perp} + {\mathbf{k}_{\perp i}}, \lambda_i\right>,
\label{a318}
\end{eqnarray}
where we consider a proton with momentum 
$P$
and spin projection $S_z$ along the $\mathbf{z}\equiv \mathbf{x}^3$ axis. 
The Fock state $n$ contains $n$
constituents, and we sum over the helicities, $\{\lambda_i\}$, 
of the constituents as well. 
The light-cone momentum fractions $x_i = k^+_i/P^+$ and ${\mathbf{
k}_{\perp i}}$ represent the relative momentum coordinates of 
constituent $i$ in Fock state $n$, whereas 
the physical momentum coordinates of constituent $i$ are
$k^+_i$ and ${\mathbf{p}_{\perp i}}
= x_i {\mathbf{P}_\perp} + {\mathbf{k}_{\perp i}}.$ 
The label $\lambda_i$ determines the helicity of a constituent 
quark or gluon along the
$\mathbf{z}$ axis. 
A free fermion constituent of mass $m_i$ is specified not only by 
its momentum components $k_i^+$, $\mathbf{k}_{\perp,i}$ 
and helicity $\lambda_i$, 
but also by its color $c_i$  and flavor $f_i$. In writing 
Eq.~(\ref{a318}) we suppress the presence of $c_i$ and $f_i$ in
the arguments of the free Fock states and light-front wave functions
for notational simplicity. Note, too, that we also 
implicitly sum over 
the constituents' colors, $\{c_i\}$, and flavors, $\{f_i\}$. 
The $n$-particle states are normalized as
\begin{equation}
\!\!
\left< n;\, p'_i{}^+, {\mathbf{p\,'}_{\perp i}}, \lambda'_i \right. \,
\left\vert n;\,
p^{~}_i{}^{\!\!+}, {\mathbf{p}_{\perp i}}, \lambda_i\right>
= \prod_{i=1}^n \!\left( 16\pi^3
  p_i^+ \delta(p'_i{}^{+} - p^{~}_i{}^{\!\!+})\
 \! \delta^{(2)}( {\mathbf{p\,'}_{\perp i}} - {\mathbf{p}_{\perp i}})\
 \! \delta_{\lambda'_i \lambda^{~}_i}\!\right) \,. \!\!\!
\label{normalize}
\end{equation}
The solutions of $ H_{LC} \ket{\psi_p} = M^2 \ket{\psi_p}$ are
independent of $P^+$ and ${\mathbf{P}_\perp}$. Thus, given the
Fock projections $ \langle n;\ x_i, {\mathbf{k}_{\perp i}},
\lambda_i |\psi_p (P^+, \mathbf{P}_\perp, S_z) \rangle$, or 
$\psi^{S_z}_{n/p}(x_i, {\mathbf{k}_{\perp i}},\lambda_i)$, 
the wave function of the proton is determined in any frame
\cite{Lepage:1980fj}. The light-front wave functions $\psi^{S_z}_{n/h}(x_i,
\mathbf{k}_{\perp i},\lambda_i)$ encode all of the bound-state quark and gluon
properties of a hadron $h$, including its momentum, spin, and flavor
correlations, in the form of universal process- and frame-independent
amplitudes.

\section{The Light-Front Representation of the Electromagnetic Form Factors}

In the case of a spin-${1/2}$ system, with exact 
eigenstate 
$|P,S_z\rangle$, the Dirac and 
Pauli form factors $F_1(q^2)$ and $F_2(q^2)$, and the 
electric dipole moment form factor $F_3(q^2)$ are defined by
\begin{eqnarray}
      \langle P', S_z^\prime | J^\mu (0) |P, S_z\rangle
       &=& \bar u(P',\lambda^\prime)\, \Big[\, F_1(q^2)\gamma^\mu +
F_2(q^2){i\over 2M}\sigma^{\mu\alpha}q_\alpha \nonumber\\ 
&& +
F_3(q^2){-1\over 2M}\sigma^{\mu\alpha}\gamma_5q_\alpha\, \Big] \, u(P,\lambda)
\ ,
\label{Drell1n}
\end{eqnarray}
where $q^\mu = (P' -P)^\mu$ and $u(P,\lambda)$ is the Dirac spinor associated
with a spin-$1/2$ state of momentum $P$ and helicity $\lambda$. 
We employ the standard light-cone frame throughout, 
so that 
$q = (q^+,q^-,\mathbf{q}_{\perp}) = (0, -q^2/P^+,
\mathbf{q}_{\perp})$ and 
$P = (P^+,P^-,\mathbf{P}_{\perp}) = (P^+, M^2/P^+, \mathbf{0}_{\perp})$, 
where $q^2=-2 P \cdot q= -\mathbf{q}_{\perp}^2$ is the square of the
momentum transferred by the photon to the system. 
We detail other pertinent conventions 
in the Appendix and note 
\begin{eqnarray}
{1 \over 2 P^+}\
{\bar u}(P',{\lambda}')\gamma^+\, u(P,\lambda)\ &=&\
\delta_{\lambda ,\, {\lambda}'}\ , \nonumber \\
{1 \over 2 P^+}\
{\bar u}(P',{\lambda}')i{\sigma^{+1}}\,(\gamma_5)\, u(P,\lambda) &=&
- \lambda\ (\lambda)\
\delta_{\lambda ,\, -{\lambda}'}\ ,
\label{tt1}\\
{1 \over 2 P^+}\
{\bar u}(P',{\lambda}')i{\sigma^{+2}}\,(\gamma_5)\, u(P,\lambda) &=&
-\ i\ (\lambda)\
\delta_{\lambda ,\, -{\lambda}'}\ .
\nonumber
\end{eqnarray}
Using Eq.~(\ref{tt1}) in conjunction with Eq.~(\ref{Drell1n}) we find 
\begin{equation} 
F_1(q^2)
=\VEV{P+q,\uparrow\left|\frac{J^+(0)}{2P^+} 
\right|P,\uparrow}
=\VEV{P+q,\downarrow\left|\frac{J^+(0)}{2P^+}
\right|P,\downarrow} 
\ ,  
\label{BD1n}
\end{equation}
\begin{equation}
{F_2(q^2)\over 2M}=
{1\over 2}\ \left[\, 
- \frac{1}{q^L}
\VEV{P+q,\uparrow\left|\frac{J^+(0)}{2P^+}\right|P,\downarrow}
+ \frac{1}{q^R}
\VEV{P+q,\downarrow\left|\frac{J^+(0)}{2P^+}\right|P,\uparrow}\, \right]
\ ,
\label{BD2n}
\end{equation}
\begin{equation}
{F_3(q^2)\over 2M}=
{i\over 2}\ \left[\, 
- \frac{1}{q^L}
\VEV{P+q,\uparrow\left|\frac{J^+(0)}{2P^+}\right|P,\downarrow}
- \frac{1}{q^R}
\VEV{P+q,\downarrow\left|\frac{J^+(0)}{2P^+}\right|P,\uparrow}\, \right]
\,,
\label{BD3n}
\end{equation}
where $\uparrow$ and $\downarrow$ denote spin states aligned parallel
and antiparallel to the $\mathbf{z}$ axis and 
$q^{{R},{L}} = q^1 \pm i q^2$. 
The Dirac and Pauli
form factors, for $q^2 \le 0$, can thus be identified from the 
helicity-conserving and helicity-flip vector-current matrix elements of
the $J^+(0)$ current in the $q^+=0$ frame~\cite{BrodskyDrell} --- 
and we find this true of $F_3(q^2)$ as well. 
The magnetic and electric dipole moments are defined in the $q^2 \to 0$
limit, namely, 
\begin{equation}
\mu=\frac{e}{2 M}\left[ F_1(0)+F_2(0) \right] \ ,\qquad
d=\frac{e}{M}F_3(0) \ ,
\label{DPmu}
\end{equation}
where $e$ is the charge and $M$ is the mass of the proton 
if we consider a spin-$1/2$ baryon 
system. Recall that $\kappa=(e/2M)F_2(0)$ is the anomalous magnetic moment. 
For leptons, such as the electron or neutrino,  
it is convenient to employ the electron mass for $M$, so that 
the magnetic moment is given in Bohr magnetons. 

Now we turn to the evaluation of the 
helicity-conserving and helicity-flip vector-current matrix elements 
in the light-front formalism. 
In the 
interaction picture, the current 
$J^\mu(0)$  is represented as a bilinear 
product of free fields, so that 
it has an elementary coupling to the 
constituent fields~\cite{DrellYan,BrodskyDrell,West}. 
The Dirac form factor can then be calculated from the
expression
\begin{equation}
F_1 (q^2) = 
\sum_{a}  \int
[{\mathrm d} x] [{\mathrm d}^2 \mathbf{k}_{\perp}]
\sum_j e_j \  
\Big[\,  
\psi^{\uparrow *}_{a}(x_i,\mathbf{k}^\prime_{\perp
i},\lambda_i) \,
\psi^\uparrow_{a} (x_i, \mathbf{k}_{\perp i},\lambda_i) \,\Big] \, , 
\label{LCch}
\end{equation} 
whereas  
the Pauli and electric dipole form factors 
are given by 
\begin{eqnarray}
&&{F_2(q^2)\over 2M} \ =\
\sum_{a}  \int
[{\mathrm d} x] [{\mathrm d}^2 \mathbf{k}_{\perp}]
\sum_j e_j \ {1\over 2}\ \times
\label{LCmu}\\
&&
\Big[\,  
-\frac{1}{q^L}
\psi^{\uparrow *}_{a}(x_i,\mathbf{k}^\prime_{\perp
i},\lambda_i) \,
\psi^\downarrow_{a} (x_i, \mathbf{k}_{\perp i},\lambda_i)
+ \frac{1}{q^R}
\psi^{\downarrow *}_{a}(x_i,\mathbf{k}^\prime_{\perp
i},\lambda_i) \,
\psi^\uparrow_{a} (x_i, \mathbf{k}_{\perp i},\lambda_i)\, \Big]
{}\ ,
\nonumber
\end{eqnarray}
\begin{eqnarray}
&&{F_3(q^2)\over 2M} \ =\
\sum_{a}  \int
[{\mathrm d} x] [{\mathrm d}^2 \mathbf{k}_{\perp}]
\sum_j e_j \ {i\over 2}\ \times
\label{LCmuF3}\\
&&
\Big[\, 
-\frac{1}{q^L}
\psi^{\uparrow *}_{a}(x_i,\mathbf{k}^\prime_{\perp
i},\lambda_i) \,
\psi^\downarrow_{a} (x_i, \mathbf{k}_{\perp i},\lambda_i)
- \frac{1}{q^R}
\psi^{\downarrow *}_{a}(x_i,\mathbf{k}^\prime_{\perp i},\lambda_i) \,
\psi^\uparrow_{a} (x_i, \mathbf{k}_{\perp i},\lambda_i)\, \Big]
{}\ . 
\nonumber
\end{eqnarray}
The summations are over all contributing Fock states $a$ and struck
constituent charges $e_j$. Here, as earlier, we refrain from including 
the constituents' color and flavor dependence in the arguments of 
the light-front 
wave functions. 
The phase-space integration is 
\begin{equation}
\int [\mathrm{d}x]\,[\mathrm{d}^2\mathbf{k}_\perp] 
\equiv \!\!\sum_{\lambda_i, c_i, f_i}\!
\left[ \,
 \prod_{i=1}^{n}
\left( \int\!\!\int {{\rm d}x_i\, {\rm d}^2 {\mathbf{k}_{\perp i}}
\over  2 (2 \pi)^3}\ \right) \right] 
16\pi^3 \delta\left(1-\sum_{i=1}^{n} x_i\right) 
\delta^{(2)}\left(\sum_{i=1}^{n} {\mathbf{k}_{\perp i}}\right)\,,
\label{phasespace}
\end{equation}
where $n$ denotes the number of constituents in Fock state $a$
and we sum over the possible $\{\lambda_i\}$, $\{c_i\}$, and $\{f_i\}$ 
in state $a$. 
The arguments of the final-state, 
light-front wave function differentiate between the struck
and spectator constituents; namely, we have~\cite{DrellYan,West}
\begin{equation}
\mathbf{k}'_{\perp j}=\mathbf{k}_{\perp j}+(1-x_j)
\mathbf{q}_{\perp}
\label{kprime1}
\end{equation}
for the struck constituent $j$ and
\begin{equation}
\mathbf{k}'_{\perp i}=\mathbf{k}_{\perp i}-x_i \mathbf{q}_{\perp}
\label{kprime2}
\end{equation}
for each spectator $i$, where $i\ne j$.
Note that because of 
the frame choice $q^+=0$, 
only diagonal ($n^\prime=n$) overlaps of the 
light-front Fock states appear~\cite{BrodskyDrell}. 

The simple expressions of Eqs.~(\ref{LCmu}) and (\ref{LCmuF3})
rely on the ability to employ the interaction
picture for the electromagnetic current and on the assumed
simple structure of the vacuum in the light-front formalism. 
Indeed, the $k^+ > 0$ constraint for massive 
particles in the light-front formalism removes all $q\bar q$
pairs from the physical vacuum. However, gluon modes, which
are massless, may possess $k^+=0$ and 
$\mathbf{k}_\perp=0$~\cite{Gribov} and can contribute,
in principle, in color-singlet combinations to the 
physical vacuum~\cite{Lenz}. 
If these contributions do not enter, then 
the free, Fock-space vacuum is also 
an eigenstate of $H_{LC}$ and the relations 
of Eqs.~(\ref{LCmu}) and (\ref{LCmuF3}) follow. 
As an example where zero modes do occur, and indeed are
essential to the description of spontaneous symmetry breaking 
and the Higgs mechanism, see Ref.~\cite{Srivastava:2002mw}. 
For a discussion of the role of zero modes in the vacuum structure
of the (chiral) Schwinger model, see 
Refs.~\cite{Srivastava:1998et,Nakawaki:1999ee}. 
It is worth noting 
that an explicit computation of the 
electron's anomalous magnetic moment, $(g-2)/2$, 
in light-front perturbation theory 
yields the expected result~\cite{BRS}: in this case, 
photon zero modes simply do not appear. 
The putative gluon zero modes are electrically
neutral, so that the electromagnetic coupling of the photon to the
constituent fields would be given 
by the quark charges regardless; the overlap formulas of 
Eqs.~(\ref{LCmu}) and (\ref{LCmuF3}) could miss a 
contribution, however, when a spectator gluon has zero 
$k^+$ and $\mathbf{k}_\perp$.

We now turn to the development of discrete symmetry transformations
in the light-front formalism, in order to ascertain the features of
the light-front wave functions needed to give rise to a nonzero
value of $F_3(q^2)$.

\section{Discrete Symmetries on the Light Front}

The development of the transformation properties of the various
fermion bilinears under $P$, $T$, and $C$ in the light-front
formalism can be made in a manner analogous to 
that of the equal-time formalism~\cite{Peskin:1995ev}.
One 
crucial difference, however, is that we invoke the
transformation properties on the 
perpendicular 
components of 
$k^\mu$ only, so that we can avoid 
transformations
such as $k^+ \leftrightarrow k^-$, or negative definite values of
$k^+$ or $k^-$. To be specific, we consider transformations on
$\mathbf{k}_\perp$ alone, so that $|\mathbf{k}_\perp|^2$, $k^-$, and $k^+$
all remain unchanged. This means that our particles will remain on
their energy shell throughout, in analogy to the on-mass-shell
condition in the equal-time formalism.

\subsection{Parity}

To implement the light-cone parity operation ${\cal P}_\perp$ we let
the spatial components of any vector $d^\mu$ transform as $d^R \to -
d^L,\,\, d^L \to - d^R $, $d^\pm\to d^\pm$. This is equivalent to
letting $d^1 \to - d^1$, with
 all other components transforming into themselves. Note that
if we do not flip $d^3$ we cannot flip the signs of both $d^1$ 
and $d^2$, as this can be realized via a continuous Lorentz
transformation from the identity. Flipping the sign of $d^1$ alone
does yield an improper Lorentz transformation, as needed, and we
would find analogous results were we to flip simply the sign of $d^2$
instead. Considering the commutator $[x_i, p_j]=\mathrm{i}
\delta_{ij}$ and $\mathbf{L}=\mathbf{r}\times \mathbf{p}$, we find
that ${\cal P}_\perp$ is a unitary 
operator and that it flips 
the spin as well. We thus realize the
parity transformation at the operator level via
\begin{eqnarray}
{\cal P}_\perp a^\lambda_{p^L,p^R} {\cal P}_\perp^\dagger
&=& \eta_a a^{-\lambda}_{-p^R,-p^L}  \,, \nonumber \\
{\cal P}_\perp b^{\lambda}_{p^L,p^R} {\cal P}_\perp^\dagger
&=& \eta_b b^{-\lambda}_{-p^R,-p^L}  \,,
\end{eqnarray}
where we suppress, here and throughout, 
possible internal indices such as color or flavor in the
fermion and antifermion annihilation operators, respectively. 
The fermion field operator $\psi(x)$ on the light-front, namely, 
\begin{equation}
\psi(x) 
\!\!=\!\! \int \frac{d {k}^+ \, d^2 \mathbf{k}_\perp }
{\sqrt{2k^+ (2\pi)^3}}
\{
a^{\lambda}_{{k}^L, {k}^R} 
u(k,\lambda) \exp(-i k\cdot x) 
+
b^{\dagger\,\lambda}_{{k}^L, {k}^R} 
v ( k,\lambda) \exp(i k\cdot x )\} 
\,,
\end{equation}
thus transforms as
\begin{eqnarray}
{\cal P}_\perp \psi(x) {\cal P}_\perp^\dagger
\!\!&=&\!\! \int \frac{d\tilde{k}^+ \, d^2 \mathbf{\tilde{k}}_\perp }
{\sqrt{2\tilde{k}^+ (2\pi)^3}}
\eta_a \sum_\lambda \{
a^{-\lambda}_{\tilde{k}^L, \tilde{k}^R} \gamma^1 \gamma_5
u(\tilde k,-\lambda) \exp(-i\tilde k\cdot (x^+,x^-, - x^R,-x^L)) \nonumber\\
&+&
b^{\dagger\,-\lambda}_{\tilde{k}^L, \tilde{k}^R} \gamma^1 \gamma_5
v (\tilde k,-\lambda) \exp(i\tilde k\cdot (x^+,x^-, - x^R,-x^L))\} 
\nonumber \\
&=& \eta_a \gamma^1 \gamma_5 \psi(x^+, x^-, - x^R, - x^L)\,,
\end{eqnarray}
where we note
${\tilde k}^\mu \equiv (k^+, k^-, - k^R, - k^L)$,
$u(k,\lambda)=\gamma^1 \gamma^5 u(\tilde k,\lambda)$, 
$v(k,\lambda)=\gamma^1 \gamma^5 v(\tilde k,\lambda)$, 
and $\eta_b^\ast = \eta_a$.
With 
\begin{equation}
{\cal P}_\perp \psi^\dagger(x) {\cal P}_\perp^\dagger
= \eta_a^\ast \gamma^1 \gamma_5 \psi^\dagger(x^+, x^-, - x^R, - x^L)\,, 
\end{equation}
and $|\eta_a|^2=1$, we thus conclude that 
\begin{eqnarray}
{\cal P}_\perp {\bar \psi}\psi(x) {\cal P}_\perp^\dagger
&=& {\bar\psi} \psi(x^+, x^-, - x^R, - x^L) \,,\\
{\cal P}_\perp i{\bar \psi}\gamma_5\psi(x) {\cal P}_\perp^\dagger
&=& - i{\bar\psi} \gamma_5 \psi(x^+, x^-, - x^R, - x^L) \,,\\
{\cal P}_\perp {\bar \psi}\gamma^\mu \psi(x) {\cal P}_\perp^\dagger
&=& \xi^\mu {\bar\psi} \gamma^\mu \psi(x^+, x^-, - x^R, - x^L) \,,
\label{Pvector}\\
{\cal P}_\perp {\bar \psi}\gamma^\mu \gamma_5\psi(x) {\cal P}_\perp^\dagger
&=& -\xi^\mu {\bar\psi} \gamma^\mu \gamma_5
\psi(x^+, x^-, - x^R, - x^L) \,,
\end{eqnarray}
where $\xi^\mu=-1$ for $\mu=1$ and $\xi^\mu=+1$ for $\mu\ne 1$. 
Repeated indices in $\mu$ are not summed. Note
that the determined vector and
axial-vector transformations are analogous to that of the equal-time
case. Moreover, we have
\begin{eqnarray}
{\cal P}_\perp {\bar \psi}\sigma^{\mu \nu}\psi(x) {\cal P}_\perp^\dagger
&=& \eta^{\mu\nu} {\bar\psi} \sigma^{\mu \nu}
\psi(x^+, x^-, - x^R, - x^L) \,,\\
{\cal P}_\perp {\bar \psi}\sigma^{\mu \nu}\gamma_5\psi(x)
{\cal P}_\perp^\dagger
&=& -\eta^{\mu\nu} {\bar\psi} \sigma^{\mu \nu} \gamma_5
\psi(x^+, x^-, - x^R, - x^L) \,,
\end{eqnarray}
where $\eta^{\mu\nu} = \xi^\mu \xi^\nu$ and repeated indices
in $\mu$ and $\nu$ are not summed. These transformations also 
parallel those found 
in the equal-time 
formalism. Applying these transformation properties to the matrix
elements which yield $F_2$ and $F_3$, in specific 
Eq.~(\ref{Drell1n}), as such are shared by the matrix elements of
the Dirac spinors, 
we see that $F_2$ is even and $F_3$ is odd
under ${\cal P}_\perp$. Turning to the explicit forms of
Eqs.(\ref{BD2n},\ref{BD3n}), we see that since $\lambda \to -
\lambda$, $q^R \to - q^L$, and $q^L \to - q^R$ under ${\cal
P}_\perp$, that if
\begin{eqnarray}
-\frac{1}{q^L}
\langle P+q, \uparrow | J^+(0) | P, \downarrow \rangle
&\stackrel{{\cal P}_\perp}{\to}
& \frac{1}{q^R}
\langle P+ q , \downarrow | J^+(0) | P, \uparrow \rangle  \,,
\label{Pfirstme} \\
\frac{1}{q^R} \langle P+q, \downarrow | J^+(0) | P, \uparrow \rangle
&\stackrel{{\cal P}_\perp}{\to}
& -\frac{1}{q^L}
\langle P+ q , \uparrow | J^+(0) | P, \downarrow \rangle \,,
\label{Psecondme}
\end{eqnarray}
we can conclude here as well
that $F_2$ is even and $F_3$ is odd under ${\cal P}_\perp$,
precisely as desired. Since the form factors are functions
of $q^2$ only, we note that the matrix element in the 
left-hand side (LHS) of
Eq.~(\ref{Pfirstme}) must be proportional to $q^L$, whereas the matrix
element in the LHS of Eq.~(\ref{Psecondme}) must be proportional to
$q^R$. Indeed, if 
\begin{equation}
-\frac{1}{q^L}
\langle P+q, \uparrow | J^+(0) | P, \downarrow \rangle
=
\frac{1}{q^R}
\langle P+  q , \downarrow | J^+(0) | P, \uparrow \rangle
\label{pcons}
\end{equation}
is also satisfied, 
then $F_3=0$ and ${\cal P}_\perp$ is a ``good''
symmetry.

Now let us consider the transformation properties of the 
light-front wave functions in
greater detail, as matrix elements of 
these quantities give rise to $F_3(q^2)$, 
a P-odd, T-odd observable. 
To summarize our earlier
discussion, the action of ${\cal P}_\perp$ is such that it transforms
the matrix elements 
entering $F_2(q^2)$ and $F_3(q^2)$ as per
Eqs.~(\ref{Pfirstme},\ref{Psecondme}). 
At the level of the wave functions themselves, we have
\begin{eqnarray}
\psi^\downarrow_{a} (\mathbf{k}_{\perp i}, x_i, \lambda_i)
&\stackrel{{\cal P}_\perp}{\to} &
\psi^\uparrow_{a} (\tilde \mathbf{k}_{\perp i},x_i,-\lambda_i) \,,
\label{lfwfP1} \\
\psi^\uparrow_{a} (\mathbf{k}_{\perp i}, x_i, \lambda_i)
&\stackrel{{\cal P}_\perp}{\to} &
\psi^\downarrow_{a} (\tilde \mathbf{k}_{\perp i},x_i,-\lambda_i) \,,
\label{lfwfP2}
\end{eqnarray}
with $\tilde \mathbf{k}_{\perp i} = (-k_i^1,k_i^2)$. 
These transformation
properties are consistent with those in
Eqs.~(\ref{Pfirstme},\ref{Psecondme}). 
We have suppressed 
the introduction of an overall phase factor as it is without physical 
relevance. 
Moreover, if 
$\psi^\downarrow(\mathbf{k}_{\perp i},
x_i,\lambda_i) = \psi^\uparrow(\tilde \mathbf{k}_{\perp i}, x_i,-\lambda_i)$
then Eq.~(\ref{pcons}) follows as well and $F_3(q^2)$ vanishes.
Thus to realize a nonzero  value of 
$F_3(q^2)$, we must have light-front wave functions which satisfy 
$\psi^\downarrow(\mathbf{k}_{\perp i},
x_i,\lambda_i) \ne 
\psi^\uparrow(\tilde \mathbf{k}_{\perp i}, x_i,-\lambda_i)$. 

\subsection{Time-Reversal}

In order to implement the light-cone time-reversal operation ${\cal T}_\perp$
we let the spatial components of any {\it momentum} vector transform
as $q^R \to - q^L\,,\, q^L \to - q^R $, so that $q^\mu \to (q^+,
q^-, -q^1, q^2)$. This implies, ultimately, that the position vector
under ${\cal T}_\perp$ transforms as $x^\mu \to (-x^+, -x^-, x^1,
-x^2)$, or $x^\mu \to (-x^+, -x^-, x^R, x^L) $. 
We term the transformation
time-reversal,  since $x^0$ does flip its sign, even though other coordinates
flip sign as well. Our construction of ${\cal T}_\perp$ is tied to
that of ${\cal P}_\perp$, so that we can conserve ${\cal C}{\cal
P}_\perp {\cal T}_\perp$, where ${\cal C}$ is the charge-conjugation
operator in the light-front formalism. 
 Moreover, our choice of ${\cal T}_\perp$ yields a 
nonorthochronous operator; the ${\cal T}_\perp$ transformation should not be
connected by a continuous Lorentz transformation to the identity.
Considering the commutator $[x_i, p_j]=\mathrm{i} \delta_{ij}$ and
$\mathbf{L}=\mathbf{r}\times \mathbf{p}$, we find that ${\cal
T}_\perp$ is antiunitary as expected but that it does {\it not}
flip the spin. We thus realize the time-reversal transformation at
the operator level via
\begin{eqnarray}
{\cal T}_\perp a^\lambda_{p^L,p^R} {\cal T}_\perp^\dagger
&=& \tilde{\eta}_a a^{\lambda}_{-p^R,-p^L} \,, \\
{\cal T}_\perp b^{\lambda}_{p^L,p^R} {\cal T}_\perp^\dagger
&=& \tilde{\eta}_b b^{\lambda}_{-p^R,-p^L}  \,,
\end{eqnarray}
so that the fermion field operator $\psi(x)$ transforms as 
\begin{eqnarray}
{\cal T}_\perp \psi(x) {\cal T}_\perp^\dagger
&=& \int \frac{d\tilde{k}^+ \, d^2 \mathbf{\tilde{k}}_\perp }
{\sqrt{2\tilde{k}^+ (2\pi)^3}}
{\tilde \eta}_a \sum_\lambda \{
a^{\lambda}_{\tilde{k}^L, \tilde{k}^R} \sigma^{12}
u(\tilde k,\lambda) \exp(-i\tilde k\cdot (-x^+,-x^-, x^R,x^L)) \nonumber \\
&+&
b^{\dagger\,\lambda}_{\tilde{k}^L, \tilde{k}^R} \sigma^{12}
v (\tilde k,\lambda) \exp(i\tilde k\cdot (-x^+,-x^-, x^R, x^L))\} \,
\nonumber \\
&=& \tilde \eta_a \sigma^{12} \psi(-x^+, -x^-,  x^R,  x^L) \,,
\end{eqnarray}
where we note that
$\tilde k \equiv (k^+, k^-, - k^R, - k^L)$, 
$u(k,\lambda)=\sigma^{12} u(\tilde k,\lambda)$, 
$v(k,\lambda)=\sigma^{12} v(\tilde k,\lambda)$, 
and ${\tilde\eta}_b^\ast = -{\tilde \eta}_a$.
With 
\begin{equation}
{\cal T}_\perp \psi^\dagger(x) {\cal T}_\perp^\dagger
= {\tilde \eta}_a^\ast \sigma^{12} \psi^\dagger(-x^+,-x^-, x^R,  x^L)\,
\end{equation}
and $|\tilde\eta_a|^2=1$, 
we thus conclude that 
\begin{eqnarray}
{\cal T}_\perp {\bar \psi}\psi(x) {\cal T}_\perp^\dagger
&=& {\bar\psi} \psi(-x^+, -x^-,  x^R,  x^L) \,,\\
{\cal T}_\perp i{\bar \psi}\gamma_5\psi(x) {\cal T}_\perp^\dagger
&=& - i{\bar\psi} \gamma_5 \psi(-x^+, -x^-,  x^R,  x^L) \,,\\
{\cal T}_\perp {\bar \psi}\gamma^\mu \psi(x) {\cal T}_\perp^\dagger
&=& \xi^\mu {\bar\psi} \gamma^\mu \psi(-x^+, -x^-,  x^R,  x^L) \,,\\
{\cal T}_\perp {\bar \psi}\gamma^\mu \gamma_5\psi(x) {\cal T}_\perp^\dagger
&=& \xi^\mu {\bar\psi} \gamma^\mu \gamma_5
\psi(-x^+, -x^-,  x^R,  x^L) \,,
\end{eqnarray}
where $\xi^\mu=-1$ for $\mu=1$ and $\xi^\mu=+1$ for $\mu\ne 1$. 
Repeated indices in $\mu$ are not summed. The transformations
found parallel that of the equal-time case. Moreover, 
\begin{eqnarray}
{\cal T}_\perp {\bar \psi}\sigma^{\mu \nu}\psi(x) {\cal T}_\perp^\dagger
&=& -\eta^{\mu\nu} {\bar\psi} \sigma^{\mu \nu}
\psi(-x^+, -x^-,  x^R,  x^L) \,,\\
{\cal T}_\perp {\bar \psi}\sigma^{\mu \nu}\gamma_5\psi(x)
{\cal T}_\perp^\dagger
&=& -\eta^{\mu\nu} {\bar\psi} \sigma^{\mu \nu} \gamma_5
\psi(-x^+, -x^-, x^R, x^L) \,,
\end{eqnarray}
where $\eta^{\mu\nu} = \xi^\mu \xi^\nu$ and, once again,
we do not sum repeated indices 
in $\mu$ and $\nu$. These transformations 
also parallel
those of the equal-time formalism. 
Since $q^\mu$ and $i q^\mu$ yield a $\xi^\mu$ and $-\xi^\mu$
under ${\cal T}_\perp$, respectively, we thus see upon applying
${\cal T}_\perp$ to Eq.~(\ref{Drell1n}) 
that $\hbox{Re} F_2$ is 
even and $\hbox{Re} F_3$ is odd, whereas 
$\hbox{Im} F_2$ is odd and $\hbox{Im} F_3$ is even. 
Applying these
transformation properties to 
the explicit forms in Eqs.(\ref{BD2n},\ref{BD3n}) for $F_2(q^2)$
and $F_3(q^2)$, as such are shared by the matrix elements of
Dirac spinors, 
we see, since $\lambda \to \lambda$, $q^R \to - q^R$, and $q^L \to -
q^L$ under ${\cal T}_\perp$, 
and ${\cal T}_\perp$
is antiunitary, that if 
\begin{eqnarray}
\langle P+q, \uparrow | J^+(0) | P, \downarrow \rangle
&\stackrel{{\cal T}_\perp}{\to}
& ( \langle P +  
\tilde q , \uparrow | J^+(0) | P, \downarrow \rangle )^\ast \, \nonumber \\
&& = 
 - \langle P+ q , \uparrow | J^+(0) | P, \downarrow \rangle  \,, 
\label{Tfirstme} \\
\langle P+q, \downarrow | J^+(0) | P, \uparrow \rangle
&\stackrel{{\cal T}_\perp}{\to}
& ( \langle P+ \tilde q , \downarrow | J^+(0) | P, \uparrow \rangle)^\ast 
\, \nonumber \\
&& 
=- \langle P+  q , \downarrow | J^+(0) | P, \uparrow \rangle \,,
\label{Tsecondme}
\end{eqnarray}
with $\tilde q = (q^+,q^-,\tilde {\mathbf q}_\perp)$
and $\tilde {\mathbf q}_\perp = (-q^1,q^2)$, 
then $\hbox{Re}(F_2)$ and $\hbox{Im}(F_3)$ 
are even 
and $\hbox{Re}(F_3)$ and $\hbox{Im}(F_2)$ 
are odd under ${\cal T}_\perp$, precisely as desired. 
As we shall see, 
the equalities emerge naturally if 
a unit of orbital angular momentum distinguishes the 
spin-up and spin-down 
light-front wave functions for 
fixed $\lambda_i$, that is, for 
fixed bachelor quark helicity in a spin-1/2 $q(qq)_0$ wave function 
--- if the 
light-front wave functions are assumed to be otherwise real. With this, 
we see that both $\hbox{Im}(F_2)$  and $\hbox{Re}(F_3)$ vanish. In order
to realize nonzero ${\cal T}_\perp$-odd observables, we will have
to allow the light-front wave functions to have additional complex
contributions. 

Now let us consider the transformation properties of the 
light-front wave functions under 
${\cal T}_\perp$ in detail. 
At the level of the 
light-front wave functions themselves, we have
\begin{eqnarray}
\psi^\downarrow_{a} (\mathbf{k}_{\perp i}, x_i, \lambda_i)
&\stackrel{{\cal T}_\perp}{\to} &
\psi^{\downarrow\,\ast}_{a} (\mathbf{\tilde k}_{\perp i},x_i,\lambda_i)
\,,
\label{genlfwfT1} \\
\psi^\uparrow_{a} (\mathbf{k}_{\perp i}, x_i, \lambda_i)
&\stackrel{{\cal T}_\perp}{\to} &
\psi^{\uparrow\,\ast}_{a} ( \mathbf{\tilde k}_{\perp i},x_i,\lambda_i) \,,
\label{genlfwfT2}
\end{eqnarray}
with $\tilde \mathbf{k}_{\perp i} = (-k^1_i,k^2_i)$. 
We suppress
the introduction of an overall phase factor as it is without physical 
impact. 
Under the assumptions which lead to 
Eqs.~(\ref{Tfirstme},\ref{Tsecondme}), 
Eqs.~(\ref{genlfwfT1},\ref{genlfwfT2}) yield 
\begin{eqnarray}
\psi_a^{\uparrow\,\ast} (\mathbf{k}_{\perp i}^\prime, x_i, \lambda_i)
\psi_a^\downarrow (\mathbf{k}_{\perp i}, x_i, \lambda_i)
&\stackrel{{\cal T}_\perp}{\to} &
-\psi_a^{\uparrow\,\ast} (\mathbf{k}_{\perp i}^\prime,x_i,\lambda_i)
\psi_a^{\downarrow} (\mathbf{k}_{\perp i},x_i,\lambda_i)
\,,
\label{lfwfT1} \\
\psi_a^{\downarrow\,\ast} (\mathbf{k}_{\perp i}^\prime, x_i, \lambda_i)
\psi_a^\uparrow (\mathbf{k}_{\perp i}, x_i, \lambda_i)
&\stackrel{{\cal T}_\perp}{\to} &
-\psi_a^{\downarrow\,\ast} ( \mathbf{k}_{\perp i}^\prime,x_i,\lambda_i)
\psi_a^{\uparrow} ( \mathbf{k}_{\perp i},x_i,\lambda_i) \,,
\label{lfwfT2}
\end{eqnarray}
for any $\lambda_i$, and 
are thus consistent with the transformations of 
Eqs.~(\ref{Tfirstme},\ref{Tsecondme}). 
Allowing the light-front wave functions to have additional complex 
contributions will enable a nonzero value of $F_3(q^2)$, as we shall
discuss in Sec.~\ref{oddlfwf}.

\subsection{Charge Conjugation}

We realize the light-front, charge-conjugation transformation
at the operator level via
\begin{eqnarray}
{\cal C} a^\lambda_{p^L,\,p^R} {\cal C}^\dagger
&=& \eta_a b^{\lambda}_{p^L,\,p^R} , \\
{\cal C} b^{\lambda}_{p^L,\,p^R} {\cal C}^\dagger 
&=& \eta_b a^{\lambda}_{p^L,\,p^R}  \,,
\end{eqnarray}
precisely as in the equal-time formalism~\cite{Peskin:1995ev}.
Indeed, we conclude in this case, as well, that
\begin{equation}
{\cal C} \psi(x) {\cal C}^\dagger = -i \gamma^2 \psi^\ast(x)\,.
\end{equation}
Note that the action of ${\cal C}$ carries $\psi \to \psi^*$, though
${\cal C}$ is a unitary operator. 
Nevertheless, 
with ${\cal C}$, ${\cal T_\perp}$, and ${\cal P_\perp}$ as we have
defined them, all scalar fermion bilinears are invariant under 
${\cal C}{\cal P_\perp}{\cal T_\perp}$, as they ought be. 
We note, e.g., that ${\bar 
\psi}\gamma^\mu \psi$, ${\bar \psi}\sigma^{\mu \nu}\psi$, and ${\bar
\psi}\sigma^{\mu \nu}\gamma_5\psi$ all yield $-1$ under ${\cal C}$, so 
that these operators yield $-1$, $+1$ , and $-1$, respectively, 
under the combined action of ${\cal C}{\cal P_\perp}{\cal T_\perp}$. 
If we employ the derivative operator $\partial^\mu$, which transforms
with a $-1$ under ${\cal C}{\cal P_\perp}{\cal T_\perp}$, to generate 
a scalar bilinear from these operators, we do indeed find that the
only nonvanishing operator transforms with a $+1$ under 
${\cal C}{\cal P_\perp}{\cal T_\perp}$. 

Let us now turn to the
transformation properties of Eq.~(\ref{Drell1n}). We note 
\begin{eqnarray}
{\cal C} \bar u(k^\prime,\lambda^\prime) \gamma^\mu 
u(k,\lambda){\cal C}^\dagger &=& 
\bar v (k,\lambda) \gamma^\mu v(k^\prime,\lambda^\prime) \,,\\
{\cal C} \bar u(k^\prime,\lambda^\prime)
\sigma^{\mu\nu} u(k,\lambda){\cal C}^\dagger &=& 
\bar v(k,\lambda) \sigma^{\mu\nu} v(k^\prime,\lambda^\prime) \,,\\
{\cal C} \bar u(k^\prime,\lambda^\prime) 
\sigma^{\mu\nu} \gamma_5 u(k,\lambda){\cal C}^\dagger &=& 
\bar v (k,\lambda) \sigma^{\mu\nu} \gamma_5 v(k^\prime,\lambda^\prime) \,.
\end{eqnarray}
Since $\bar v(k,\lambda) \gamma^\mu v(k^\prime,\lambda^\prime) 
= \bar u(k^\prime,\lambda^\prime) \gamma^\mu u(k,\lambda)$
and $q^\mu$ transforms with a +1 under 
${\cal C}{\cal P_\perp}{\cal T_\perp}$, 
we see that the scalar 
bilinear formed by contracting Eq.~(\ref{Drell1n}) with $q^\mu$ 
does transform with a $+1$ under ${\cal C}{\cal P_\perp}{\cal T_\perp}$,
as needed. Writing the analogue of Eq.~(\ref{Drell1n}) for an 
antifermion $\bar f$, replacing $F_i(q^2)$ with 
$\tilde F_i(q^2)$,  and evaluating the spinor matrix elements, 
we find 
\begin{equation} 
\tilde F_1(q^2)
=\VEV{P+q,\uparrow\left|\frac{J^+(0)}{2P^+} 
\right|P,\uparrow}_{\bar f}
=\VEV{P+q,\downarrow\left|\frac{J^+(0)}{2P^+}
\right|P,\downarrow}_{\bar f}
\ ,  
\label{barBD1n}
\end{equation}
\begin{equation}
{\tilde F_2(q^2)\over 2M}=
{1\over 2}\ \left[\, 
- \frac{1}{q^L}
\VEV{P+q,\uparrow\left|\frac{J^+(0)}{2P^+}\right|P,\downarrow}_{\bar f}
+ \frac{1}{q^R}
\VEV{P+q,\downarrow\left|\frac{J^+(0)}{2P^+}\right|P,\uparrow}_{\bar f}
\, \right]
\ ,
\label{barBD2n}
\end{equation}
and 
\begin{equation}
{\tilde F_3(q^2)\over 2M}=
{i\over 2}\ \left[\, 
- \frac{1}{q^L}
\VEV{P+q,\uparrow\left|\frac{J^+(0)}{2P^+}\right|P,\downarrow}_{\bar f}
- \frac{1}{q^R}
\VEV{P+q,\downarrow\left|\frac{J^+(0)}{2P^+}\right|P,\uparrow}_{\bar f}\, 
\right]
\,,  
\label{barBD3n}
\end{equation}
where 
\begin{equation}
\mu= - \frac{e}{2 M}\left[ \tilde F_1(0)+ \tilde F_2(0) \right] \ ,\qquad
d=-\frac{e}{M} \tilde F_3(0) \ ,
\label{barDPmu}
\end{equation}
and the electric charge of $\bar f$ is given by 
$-e \tilde F_1(0)$. Consequently, 
we infer the transformation properties 
\begin{eqnarray}
\langle P+q, \uparrow | J^+(0) | P, \uparrow \rangle_{f}
&\stackrel{{\cal C}}{\to}
&   \langle P+ q , \uparrow | J^+(0) | P, \uparrow \rangle_{\bar f}  \,,
\\
\langle P+q, \downarrow | J^+(0) | P, \downarrow \rangle_{f}
&\stackrel{{\cal C}}{\to}
&   \langle P+  q , \downarrow | J^+(0) | P, \downarrow \rangle_{\bar f} \,,
\end{eqnarray}
and
\begin{eqnarray}
\langle P+q, \uparrow | J^+(0) | P, \downarrow \rangle_{f}
&\stackrel{{\cal C}}{\to}
&   \langle P+ q , \uparrow | J^+(0) | P, \downarrow \rangle_{\bar f}  \,,
\label{Cfirstme} \\
\langle P+q, \downarrow | J^+(0) | P, \uparrow \rangle_{f}
&\stackrel{{\cal C}}{\to}
&   \langle P+  q , \downarrow | J^+(0) | P, \uparrow \rangle_{\bar f} \,. 
\label{Csecondme}
\end{eqnarray}
The $f$ and $\bar f$ subscripts signify that the matrix
elements are computed for a composite fermion ($f$) and antifermion
($\bar f$), respectively. 
At the level of the light-front wave functions 
themselves, we thus have
\begin{eqnarray}
\psi^\downarrow_f (\mathbf{k}_\perp, x, \lambda)
&\stackrel{{\cal C}}{\to} &
\psi^{\downarrow}_{\bar f} (\mathbf{k}_\perp,x,\lambda)
\,,
\label{lfwfC1} \\
\psi^\uparrow_f (\mathbf{k}_\perp, x, \lambda)
&\stackrel{{\cal C}}{\to} &
\psi^{\uparrow}_{\bar f} ( \mathbf{k}_\perp,x,\lambda) \,.
\label{lfwfC2}
\end{eqnarray}
We suppress
the introduction of an overall phase factor as it is without physical 
impact.

To conclude this section we consider 
how the 
products of 
the light-front wave functions which yield the electromagnetic
form factors, noting Eqs.~(\ref{LCch},\ref{LCmu},\ref{LCmuF3}), 
behave under ${\cal C}{\cal P_\perp}{\cal T_\perp}$.
Using the transformations we have discussed, we have
\begin{eqnarray}
\psi^{\uparrow\,\ast}_f (\mathbf{k}_\perp^\prime, x, \lambda)
\psi^\uparrow_f (\mathbf{k}_\perp, x, \lambda)
&\stackrel{{\cal P_\perp}}{\to} &
\psi^{\downarrow\,\ast}_f (\mathbf{\tilde k}_\perp^\prime, x, -\lambda)
\psi^\downarrow_f (\mathbf{\tilde k}_\perp, x, -\lambda) \nonumber\\
&\stackrel{{\cal T_\perp}}{\to}&
\psi^{\downarrow\,\ast}_f (\mathbf{k}_\perp, x, -\lambda)
\psi^{\downarrow}_f (\mathbf{k}_\perp^\prime, x, -\lambda) \nonumber\\
&\stackrel{{\cal C}}{\to}&
 \psi^{\downarrow\,\ast}_{\bar f} (\mathbf{k}_\perp,x,-\lambda)
\psi^{\downarrow}_{\bar f} (\mathbf{k}_\perp^\prime,x,-\lambda)
\,.  
\label{lfwfCPT2}
\end{eqnarray}
The last, 
upon integrating over phase space as per Eq.~(\ref{phasespace}), 
with the  change of variable 
$\mathbf{k}_\perp = \mathbf{k}_\perp^\prime$, 
yields $\tilde F_1(q^2)$, 
Eq.~(\ref{barBD1n}). For $F_2(q^2)$ and $F_3(q^2)$ we consider 
\begin{eqnarray}
-\frac{1}{q^L}\psi^{\uparrow\,\ast}_f (\mathbf{k}_\perp^\prime, x, \lambda)
\psi^\downarrow_f (\mathbf{k}_\perp, x, \lambda)
&\stackrel{{\cal P_\perp}}{\to} &
\frac{1}{q^R}\psi^{\downarrow\,\ast}_f (\mathbf{\tilde k}_\perp^\prime, x, -\lambda)
\psi^\uparrow_f (\mathbf{\tilde k}_\perp, x, -\lambda) \nonumber\\
&\stackrel{{\cal T_\perp}}{\to} &
- \frac{1}{q^R}\psi^{\uparrow\,\ast}_f (\mathbf{k}_\perp, x, -\lambda)
\psi^\downarrow_f (\mathbf{k}_\perp^\prime, x, -\lambda) \nonumber\\
&\stackrel{{\cal C}}{\to}&
- \frac{1}{q^R}\psi^{\uparrow\,\ast}_{\bar f} (\mathbf{k}_\perp, x, -\lambda)
\psi^\downarrow_{\bar f} (\mathbf{k}_\perp^\prime, x, -\lambda) 
\nonumber \\
&& = 
 \frac{1}{q^R}\psi^{\uparrow\,\ast}_{\bar f} (\mathbf{k}_\perp^\prime, 
x, -\lambda)
\psi^\downarrow_{\bar f} (\mathbf{k}_\perp, x, -\lambda) 
\label{lfwfCPT1}
\end{eqnarray}
and 
\begin{eqnarray}
\frac{1}{q^R}\psi^{\downarrow\,\ast}_f (\mathbf{k}_\perp^\prime, x, \lambda)
\psi^\uparrow_f (\mathbf{k}_\perp, x, \lambda)
&\stackrel{{\cal P_\perp}}{\to} &
-\frac{1}{q^L}\psi^{\uparrow\,\ast}_f (\mathbf{\tilde k}_\perp^\prime, x, -\lambda)
\psi^\downarrow_f (\mathbf{\tilde k}_\perp, x, -\lambda) \nonumber\\
&\stackrel{{\cal T_\perp}}{\to} &
\frac{1}{q^L}\psi^{\downarrow\,\ast}_f (\mathbf{k}_\perp, x, -\lambda)
\psi^\uparrow_f (\mathbf{k}_\perp^\prime, x, -\lambda) \nonumber\\
&\stackrel{{\cal C}}{\to}&
\frac{1}{q^L}\psi^{\downarrow\,\ast}_{\bar f} (\mathbf{k}_\perp, x, -\lambda)
\psi^\uparrow_{\bar f} (\mathbf{k}_\perp^\prime, x, -\lambda) 
\nonumber \\ 
&& = -\frac{1}{q^L}\psi^{\downarrow\,\ast}_{\bar f} (\mathbf{k}_\perp^\prime, 
x, -\lambda)
\psi^\uparrow_{\bar f} (\mathbf{k}_\perp, x, -\lambda) 
\,,
\label{lfwfCPT3}
\end{eqnarray}
where the equalities arise from making the change of variable
$\mathbf{k}_\perp = \mathbf{k}_\perp^\prime$ in the integration
over phase space as per Eq.~(\ref{phasespace}). 
Starting with Eqs.~(\ref{BD2n},\ref{BD3n})
we find, under ${\cal C}{\cal P}_\perp {\cal T}_\perp$, 
these last expressions will give rise to $\tilde F_2(q^2)$
and $\tilde F_3(q^2)$, 
Eqs.~(\ref{barBD2n},\ref{barBD3n}), so that we see
explicitly that $F_2(q^2)$ and $F_3(q^2)$ --- as well as 
$F_1(q^2)$ --- yield $+1$ under 
${\cal C} {\cal P}_\perp {\cal T}_\perp$, 
 at the level of the light-front
wave functions, 
as consistent with the
transformation properties of the original 
fermion bilinears. 
This concludes our discussion of discrete symmetry transformations
on the light front.

\section{Light-Front Wave Functions for ${\cal T}_\perp$-Odd 
and ${\cal P}_\perp$-Odd Observables}
\label{oddlfwf}

In this section we develop simple light-front wave functions of the
nucleon which 
are compatible with a nonzero electric dipole moment. 
To begin, we consider a quark--scalar-diquark model of the 
nucleon, $q(qq)_0$, patterned after the interaction of a fermion
and a neutral scalar in Yukawa theory~\cite{Brodsky:2000ii}. 
This model has proved useful in the analysis of single-spin
asymmetries in semi-inclusive, deeply inelastic 
scattering~\cite{Brodsky:2002cx}. In this 
model, the nucleon light-front wave function has two particles:
a quark and a scalar diquark, so that the $J^z = + {1\over 2}$ 
nucleon wave function is taken to be of form 
\begin{eqnarray}
&&\left|\Psi^{\uparrow}_{\rm q(qq)_0}(P^+=1, \mathbf{ P}_\perp = 
\mathbf{0}_\perp)\right>
\label{sn1}\\
&=&
\int\frac{{\mathrm d} x \, {\mathrm d}^2
           \mathbf{k}_{\perp} }{\sqrt{x(1-x)}\, 16 \pi^3}
\Big[ \
\psi^{\uparrow}_{+\frac{1}{2}} (x,\mathbf{k}_{\perp})\,
\left| \,x\, , \mathbf{k}_{\perp}\,, +\frac{1}{2}  \right>
+\psi^{\uparrow}_{-\frac{1}{2}} (x,\mathbf{k}_{\perp})\,
\left| \, x\, , \mathbf{k}_{\perp}\,,-\frac{1}{2} \right>\ \Big]\ ,
\nonumber
\end{eqnarray}
where we have labelled the free Fock states and associated
light-front wave functions with the relative 
momentum coordinates, ($x$, $\mathbf{k}_\perp$), 
and spin projection along the $\mathbf{z}$-axis, 
 $\lambda$/2, of the bachelor, or unpaired, quark. Computing 
the ${\bar u}(k^\prime, \lambda^\prime) u(k,\lambda)$ matrix element, 
we have 
\begin{equation}
\left
\{ \begin{array}{l}
\psi^{\uparrow}_{+\frac{1}{2}} (x, \mathbf{k}_{\perp})=
f(x) \varphi (x,k_\perp^2) \ ,\\
\psi^{\uparrow}_{-\frac{1}{2}} (x, \mathbf{k}_{\perp})=
-(k^1+{\mathrm i} k^2)\, g(x) \varphi (x,k_\perp^2) \ , 
\end{array}
\right. 
\label{sn2}
\end{equation}
where $f(x)$, $g(x)$, and $\varphi(x,k_\perp^2)$ 
are real, scalar functions yielding a nucleon wave function
normalized to unit probability, namely, 
\begin{equation}
\langle \Psi^{\uparrow}_{\rm q(qq)_0}(P^+=1, \mathbf{ P}_\perp = 
\mathbf{0}_\perp) | \Psi^{\uparrow}_{\rm q(qq)_0}(P^+=1, \mathbf{ P}_\perp = 
\mathbf{0}_\perp)\rangle = 1  \,.
\end{equation}
Similarly, the $J^z = - {1\over 2}$ nucleon wave function 
is given by 
\begin{eqnarray}
&&\left|\Psi^{\downarrow}_{\rm q(qq)_0}(P^+=1, \mathbf{ P}_\perp = 
\mathbf{0}_\perp)\right>
\label{sn1a}\\
&=&
\int\frac{{\mathrm d} x \, {\mathrm d}^2
           \mathbf{k}_{\perp} }{\sqrt{x(1-x)}\, 16 \pi^3}
\Big[ \
\psi^{\downarrow}_{+\frac{1}{2}} (x,\mathbf{k}_{\perp})\,
\left| \,x\, , \mathbf{k}_{\perp}\,, +\frac{1}{2}  \right>
+\psi^{\downarrow}_{-\frac{1}{2}} (x,\mathbf{k}_{\perp})\,
\left| \, x\, , \mathbf{k}_{\perp}\,,-\frac{1}{2} \right>\ \Big]\ ,
\nonumber
\end{eqnarray}
where
\begin{equation}
\left
\{ \begin{array}{l}
\psi^{\downarrow}_{+\frac{1}{2}} (x,\mathbf{k}_{\perp})=
(k^1-{\mathrm i} k^2) g(x) \varphi(x, k_\perp^2)  \ ,\\
\psi^{\downarrow}_{-\frac{1}{2}} (x,\mathbf{k}_{\perp})=
f(x) \varphi (x,k_\perp^2) \ .
\end{array}
\right.
\label{sn2a}
\end{equation}
The structure of Eqs.~(\ref{sn2},\ref{sn2a}) 
is common to that of the electron-photon Fock states of 
Ref.~\cite{BrodskyDrell}. 
The light-front wave functions
of Eqs.~(\ref{sn1},\ref{sn2},\ref{sn1a},\ref{sn2a}) satisfy 
Eq.~(\ref{pcons}), 
as well as Eqs.~(\ref{Tfirstme}) and (\ref{Tsecondme}), 
so that in this model we find $F_3(q^2)=0$ and $\hbox{Im}(F_2(q^2))=0$. 

We can generalize this model, however, so that ${\cal T}_\perp$-odd
or ${\cal P}_\perp$-odd observables no longer vanish. Indeed, if
we now include phases,  writing 
\begin{eqnarray}
&& \left
\{ \begin{array}{l}
\psi^{\uparrow}_{+\frac{1}{2}} (x,\mathbf{k}_{\perp})= 
f(x) \varphi (x,k_\perp^2) \, 
e^{i\alpha_1}\, e^{+i\beta_1} \ ,\\
\psi^{\uparrow}_{-\frac{1}{2}} (x,\mathbf{k}_{\perp})=
-(k^1+{\mathrm i} k^2)
g(x) \varphi (x,k_\perp^2) \, 
e^{i\alpha_2}\, e^{+i\beta_2} \ ,
\end{array}
\right.
\label{sn2nb} \\
&& \left
\{ \begin{array}{l}
\psi^{\downarrow}_{+\frac{1}{2}} (x,\mathbf{k}_{\perp})=
(k^1-{\mathrm i} k^2) g(x) \varphi (x,k_\perp^2) \, 
e^{i\alpha_2}\, e^{-i\beta_2} \ ,\\
\psi^{\downarrow}_{-\frac{1}{2}} (x,\mathbf{k}_{\perp})=
f(x) \varphi (x,k_\perp^2) \, 
e^{i\alpha_1}\, e^{-i\beta_1}\ , 
\end{array}
\right.
\label{sn2anb}
\end{eqnarray}
where $\alpha_1$, $\alpha_2$, $\beta_1$, and $\beta_2$ are
real constants, 
$F_3(q^2)$ and $\hbox{Im}(F_2(q^2))$ can both be nonzero.
We regard $\alpha_1$, $\alpha_2$, $\beta_1$, and $\beta_2$ 
as simple constants and not as functions of $k_\perp^2$ because we 
implicitly  assume that 
the scale at which CP is broken, $M_{CP}$, in the fundamental theory
is much larger than any 
we can access experimentally, so that $q^2 \ll M_{\rm CP}^2$. 
For an explicit example of a mechanism realizing 
this, see Ref.~\cite{Hiller:2001qg}. In certain exceptional cases, 
it may be possible to have effective, $k_\perp^2$-dependent phases. 
Suppose, e.g., two distinct mechanisms of CP violation operate
in a single Fock state $a$. In that event, assuming $\beta_1$ and
$\beta_2$ are small, one could write the nucleon light-front wave 
function, suppressing all arguments, as 
$\psi_a = \psi_{1\,a} \exp(i\beta_1) + \psi_{2\,a} \exp(i\beta_2)
\approx (\psi_{1\,a} + \psi_{2\,a})  
+ i (\psi_{1\,a} \beta_1 + \psi_{2\,a} \beta_2) 
\approx (\psi_{1\,a} + \psi_{2\,a})\exp(i\tilde \beta)$, 
where
$\tilde \beta = \tan^{-1} [(\psi_{1\,a} \beta_1 + \psi_{2\,a} \beta_2)
/(\psi_{1\,a} + \psi_{2\,a})]$. 
Here we see explicitly that if $\psi_{1\,a}$ and $\psi_{2\,a}$ 
differ in their $k_\perp^2$ dependence that 
$\tilde \beta$ will be $k_\perp^2$ dependent 
even if $\beta_1$ and $\beta_2$ are not. 

Let us consider the impact of
the specific phases we have introduced. 
Firstly, we observe that if $\beta_1\ne 0$ or $\beta_2\ne 0$
Eq.~(\ref{pcons}) no longer holds, so that $\beta_1$ and $\beta_2$
generate ${\cal P}_\perp$-odd effects. 
Secondly, if $\alpha_2 - \beta_2 - \alpha_1
- \beta_1 \ne 0$ or $\alpha_2 + \beta_2 - \alpha_1 + \beta_1 \ne 0$, 
then the equalities of 
Eqs.~(\ref{Tfirstme},\ref{Tsecondme}) 
will not follow, 
and we can recover 
nonzero ${\cal T}_\perp$-odd effects. 
We evaluate $F_2(q^2)$ and $F_3(q^2)$ with these model 
wave functions in the next section and determine that
$\alpha_1 -\alpha_2\ne 0$ gives rise to ${\cal T}_\perp$-odd and 
${\cal P}_\perp$-even observables, whereas 
$\beta_1\ne 0$ or $\beta_2\ne 0$ gives rise to ${\cal T}_\perp$-odd and 
${\cal P}_\perp$-odd observables. We remark in passing that 
$\alpha_1$ and $\alpha_2$ can also be introduced to pattern 
the phases that appear in final-state interactions and 
produce the Sivers effect~\cite{Brodsky:2002cx}. Such pseudo-time-reversal-odd
effects are not produced by fundamental sources of CP violation which
are our focus here.

\section{Relating the anomalous magnetic and electric dipole moments} 

In this section we consider the relationship between 
$F_2(q^2)$ and $F_3(q^2)$ predicated by the relations of 
Eq.~(\ref{BD2n}) and (\ref{BD3n}). 

\subsection{Quark--Scalar-Diquark Model}
\label{qdiq}

We begin by computing $F_2(q^2)$ and $F_3(q^2)$ using the 
light-front wave functions of the quark--scalar-diquark model, 
Eqs.~(\ref{sn2nb},\ref{sn2anb}).
In the following ${\cal A}$ is a function given by 
\begin{equation}
{\cal A}=\int
{ {\mathrm d} x \, {\mathrm d}^2 \mathbf{k}_{\perp} \over 16 \pi^3}
\, e\, \varphi(x,{k^\prime_{\perp}}^2) \varphi(x,k_{\perp}^2) f(x) g(x) 
\label{nn1}
\end{equation}
and $\beta= \beta_1 + \beta_2$, 
where we recall that $\varphi(x, k_\perp^2)$, $f(x)$, and $g(x)$ are real. 
The bachelor quark is given charge $e$. 
From Eq.~(\ref{LCmu}) we have
\begin{eqnarray}
&&{F_2(q^2)\over 2M} =
{e\over 2}\int { {\mathrm d} x \, {\mathrm d}^2 \mathbf{k}_{\perp} \over 16 \pi^3}
\ \times
\label{note1c}\\
&& 
\Bigl(
 {1\over -q^1+{\mathrm i}q^2}\Big[ \psi^{\uparrow *}(x,\mathbf{k}^\prime_{\perp}
,{1}) \,
\psi^\downarrow (x, \mathbf{k}_{\perp},{1})
+\psi^{\uparrow *}(x,\mathbf{k}^\prime_{\perp},-{1}) \,
\psi^\downarrow (x, \mathbf{k}_{\perp},-{1})\Big] \nonumber
\nonumber\\
&&\qquad +\ {1\over q^1+{\mathrm i}q^2}
\Big[ \psi^{\downarrow *}(x,\mathbf{k}^\prime_{\perp}
,{1}) \,
\psi^\uparrow (x, \mathbf{k}_{\perp},{1})
+\psi^{\downarrow *}(x,\mathbf{k}^\prime_{\perp},-{1}) \,
\psi^\uparrow (x, \mathbf{k}_{\perp},-{1})\Big] \Bigr) \,.
\nonumber
\end{eqnarray}
Doing the $d^2{\mathbf k}_\perp$ integral, we note that the terms in
${\mathbf k}_\perp$ will vanish unless 
${\mathbf k}_\perp \parallel {\mathbf q}_\perp$,
so that we have
\begin{equation}
\frac{F_2(q^2)}{2M}= {\cal A}\cos\beta \left[(1-x)
\exp(i(\alpha_1 - \alpha_2)) + 2 i \sin(\alpha_1- \alpha_2)\right] \,.
\label{F2res}
\end{equation}
From (\ref{LCmuF3}) we have
\begin{eqnarray}
&&{F_3(q^2)\over 2M} ={i\,e\over 2}\int
{ {\mathrm d} x \, {\mathrm d}^2 \mathbf{k}_{\perp} \over 16 \pi^3} \ \times
\label{note1b}\\
&& \Bigl(
{1\over -q^1+{\mathrm i}q^2}\Big[ \psi^{\uparrow *}(x,\mathbf{k}^\prime_{\perp}
,{1}) \,
\psi^\downarrow (x, \mathbf{k}_{\perp},{1})
+\psi^{\uparrow *}(x,\mathbf{k}^\prime_{\perp},-{1}) \,
\psi^\downarrow (x, \mathbf{k}_{\perp},-{1})\Big]
\nonumber\\
&&\qquad -\ {1\over q^1+{\mathrm i}q^2}
\Big[ \psi^{\downarrow *}(x,\mathbf{k}^\prime_{\perp}
,{1}) \,
\psi^\uparrow (x, \mathbf{k}_{\perp},{1})
+\psi^{\downarrow *}(x,\mathbf{k}^\prime_{\perp},-{1}) \,
\psi^\uparrow (x, \mathbf{k}_{\perp},-{1})\Big] \Bigr) \,.
\nonumber
\end{eqnarray}
In doing the $d^2{\mathbf k}_\perp$ integral, we once again 
note that the terms in
${\mathbf k}_\perp$ will vanish unless 
${\mathbf k}_\perp \parallel {\mathbf q}_\perp$, so that 
we have
\begin{equation}
\frac{F_3(q^2)}{2M}= {\cal A} \sin\beta \left[(1-x) \exp(i(\alpha_1 - \alpha_2))
+ 2 i \sin(\alpha_1- \alpha_2)\right] \,.
\label{F3res}
\end{equation}
Comparing Eqs.~(\ref{F2res}) and (\ref{F3res}), we can elucidate
the impact of the phases we have introduced. 
For example, if $\alpha_1 \ne \alpha_2$ with 
$\beta_1=\beta_2=0$, we can have $\hbox{Im}(F_2(q^2))\ne 0$ but 
$F_3(q^2)=0$. Alternatively, 
if $\alpha_1= \alpha_2$ and $\beta_1, \beta_2 \ne 0$, 
then $\hbox{Im}(F_2(q^2))= 0$ but $F_3(q^2) \ne 0$. 
Finally if $\alpha_1 \ne \alpha_2$ and 
$\beta_1, \beta_2 \ne 0$,  then both 
$\hbox{Im}(F_2(q^2))\ne 0$ and $F_3(q^2) \ne 0$. 
It is remarkable that $F_2(q^2)$ and $F_3(q^2)$ differ
only in their explicit dependence in $\beta$. We find, in specific, that
\begin{equation}
F_3(q^2) = (\tan\beta) F_2 (q^2) \,.
\end{equation}
We proceed to examine how this relation emerges generally and
its consequences for constraints on models of CP violation. 

\subsection{General Relation}

We now consider the relationship 
between the $F_2(q^2)$ and $F_3(q^2)$ form factors
on general grounds. We will realize this by writing the light-front
wave function of
the nucleon in Fock component $a$ as
\begin{equation}
\left
\{ \begin{array}{l}
\psi^{\uparrow}_a (x_i,\mathbf{k}_{\perp\,i},\lambda_i)=
\phi^{\uparrow}_a (x_i,\mathbf{k}_{\perp\,i},\lambda_i)
e^{+i\beta_a/2}
\ ,\\
\psi^{\downarrow}_a (x_i,\mathbf{k}_{\perp\,i},\lambda_i)=
\phi^{\downarrow}_a (x_i,\mathbf{k}_{\perp\,i},\lambda_i)
e^{-i\beta_a/2} \,,
\end{array}
\right.
\label{gen2nb}
\end{equation}
where we have explicitly pulled out the ${\cal P}_\perp$- and 
${\cal T}_\perp$-violating 
parameter $\beta_a$, which we have allowed to depend on
the Fock state $a$. The remaining function
$\phi^{\uparrow}_a (x_i,\mathbf{k}_{\perp\,i},\lambda_i)$ explicitly
satisfies Eq.~(\ref{pcons}) and the
equalities of Eqs.~(\ref{Tfirstme},\ref{Tsecondme}) 
in itself. We emphasize that the parametrization of 
Eq.~(\ref{gen2nb}) is a unique and general way of introducing
CP-violation, as 
realized from a local, Lorentz-invariant 
quantum field theory with a Hermitian Hamiltonian. 
It is not specific to the Standard Model. 
With Eq.~(\ref{gen2nb})
we thus find
\begin{eqnarray}
\frac{F_2(q^2)}{2M} &=&  \sum_a \cos(\beta_a) \Xi_a \,, \\
\frac{F_3(q^2)}{2M} &=&  \sum_a \sin(\beta_a) \Xi_a \,,
\end{eqnarray}
where
\begin{equation}
\Xi_a = \int {\left[ {\mathrm d} x \right] 
\left[{\mathrm d}^2 \mathbf{k}_{\perp} \right]
\over 16 \pi^3}
\sum_j e_j \frac{1}{-q^1 + i q^2}
\Big[ \phi^{\uparrow *}_a(x_i,\mathbf{k}^\prime_{\perp\,i}
,\lambda_i) \,
\phi^\downarrow_a (x_i, \mathbf{k}_{\perp\,i},\lambda_i) \Big] \,.
\end{equation}
Thus for a particular Fock component we can write
\begin{equation}
[F_3(q^2)]_a = \tan\beta_a [F_2 (q^2)]_a \,,
\label{genrel}
\end{equation}
where the notation $[F]_a$ denotes the contribution to the form factor 
from Fock component $a$. At $q^2=0$, this becomes
\begin{equation}
d_a = 2\kappa_a \tan \beta_a \,,  
\label{central}
\end{equation}
which is the central result of our paper. 
As $\beta_a$ is ${\cal P}_\perp$- 
and ${\cal T}_\perp$-violating, we can assume it to be small, to write
\begin{equation}
[F_3(q^2)]_a = \beta_a [F_2 (q^2)]_a \,,
\label{betasmall}
\end{equation}
or
\begin{equation}
d_a = 2\kappa_a \beta_a \,. 
\end{equation}
We now proceed to consider how such connections can constrain 
theoretical predictions of $F_3(q^2)$ and hence impact bounds
on CP-violating parameters.

\section{Consequences for Models of CP-Violation}

The relation we have written, namely Eq.~(\ref{genrel}), must hold
irrespective of the possible sources of CP violation we consider: it
holds both in and beyond the Standard Model. Moreover, it is
appropriate to {\it any} spin-1/2 system, be it nucleon or lepton. 
That such a relation exists for charged leptons has been recognized by
Feng, Matchev, and Shadmi~\cite{Feng:2001sq} in supersymmetric models;
in fact, as we have shown, the anomalous magnetic moment, $g-2$, and
the electric dipole moment of the charged leptons are tied in any model.
We have extended this notion to 
neutral leptons, such as the neutrino, and to composite
spin-$1/2$ systems, such as the nucleon, as well. 

In the case of the nucleon, the 
empirical values of the anomalous magnetic moments are remarkably
well known, and we have seen that 
the essential hadronic matrix elements are also common to 
the calculation of the electric dipole moment, $d$. 
Generally, a diagrammatic calculation of an observable 
such as $d$ can be interpreted in terms of the contributions to $d$ 
from states in a Fock expansion. We need only determine the
contributions of 
the intermediate states of the system, associated with   
all possible light-cone-time-ordered graphs~\cite{Lepage:1980fj}, 
to which the photon couples. The contribution of a particular 
Fock state can thus be realized from the sum of many Feynman graphs. 
In principle, one could  compute the light-front wave functions
$\psi_{a/N}^{S_z}(x_i, \mathbf{k}_{\perp i},\lambda_i)$~\cite{Pauli:1985ps} 
and evaluate the requisite matrix elements directly, though the former
has not yet been realized in QCD. 
Fortunately, in certain models, the determination of 
the Fock state contributions, as well as of their sum, 
becomes greatly simplified. 
For example, if we were to make a
constituent quark model of the light-front wave functions,
such as the $q(qq)_0$ model of Sections~\ref{oddlfwf} and
\ref{qdiq}, 
we would be able to simplify our
relation still further to write, for small $\beta$, 
\begin{equation}
F_3(q^2) = \beta F_2 (q^2)  \,.
\end{equation}
In such a model we can estimate 
$d$ directly using the empirical
anomalous magnetic moment of the nucleon. Noting 
$\kappa^n = -1.91$ and $\kappa^p = 1.79$ (in units of $\mu_N$), we thus estimate that
\begin{equation}
d^n \sim 
- e \beta^n (2\cdot 10^{-14}\, \hbox{cm}) \,,\quad 
d^p \sim + e \beta^p (2\cdot 10^{-14}\, \hbox{cm}) \,,
\label{start}
\end{equation}
where $2 \cdot 10^{-14}$ cm is the proton mass in cm. The
current empirical bound on 
$d^n$, $|d^n| < 6.3 \cdot 10^{-26}$ e-cm~\cite{edmlimitn}, 
implies, in our simple picture, 
that $|\beta^n| < 3\cdot 10^{-12}$. If $\beta^n \sim \beta^p$, then we 
predict $d^p\sim -d^n$ and thus that 
the isoscalar electric dipole moment is small, namely, 
$|d^n + d^p| \ll |d^n -d^p|$, 
just as the empirical 
isoscalar anomalous magnetic moment is small, 
$|\kappa^n + \kappa^p| = 0.12 \ll |\kappa^n - \kappa^p| = 3.70$. 
For reference, 
we note the empirical bound on $d^p$, 
$|d^p| < 5.4\cdot 10^{-24}$ e-cm~\cite{Dmitriev:2003sc}. 
Although not apparent in this simple model, it is possible to connect 
the ${\cal T}_\perp$- and ${\cal P}_\perp$-odd parameter $\beta$
to fundamental sources of CP violation in a realistic way; we shall now 
explore this possibility. 

\begin{figure}[htb]
\begin{center}
\epsfig{file=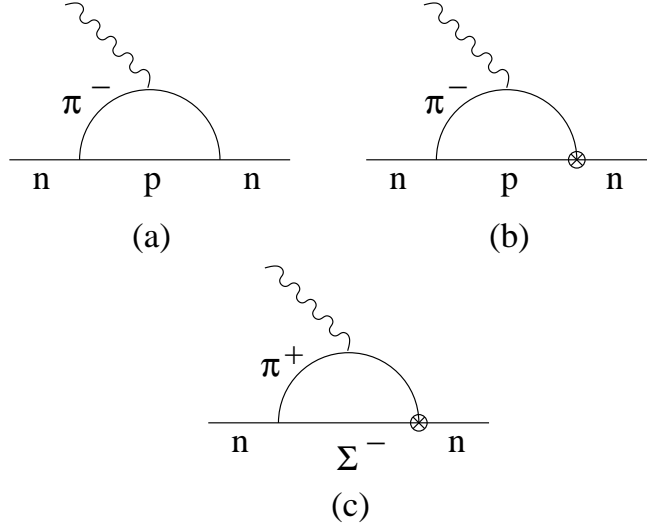, height = 7cm}
\end{center}
\centerline{\parbox{12cm}{\caption{
a) Feynman diagram for an one-loop contribution 
to the 
anomalous magnetic moment of the neutron in
heavy-baryon chiral perturbation theory (HBCHPT). 
b) Feynman diagram for an one-loop contribution 
to the electric dipole moment of the neutron 
from strong interaction CP violation in HBCHPT, where 
\protect{$\otimes$}
denotes a CP-violating vertex. 
c) Feynman diagram for an one-loop contribution 
to the electric dipole moment of the neutron 
from weak interaction 
CP violation in HBCHPT, as per the CKM mechanism of 
the Standard Model. 
\label{fig:edmloop}
}}}
\end{figure}

As an explicit example, we consider strong-interaction CP violation 
via a QCD $\theta$-term and adopt the chiral Lagrangian framework of
Refs.~\cite{Baluni,Crewther:1979pi,Pich:1991fq,Borasoy:2000pq,Hockings:2005cn} 
to estimate $d^n$ and $d^p$. 
In specific, we 
assume as in Ref.~\cite{Crewther:1979pi} that diagram b) of 
Fig.~\ref{fig:edmloop}, 
realized in terms of the meson and baryon degrees of freedom operative
in chiral effective theories at low energies, %$q^2 \ll M^2$,  
along with its counterpart containing a CP-violating
interaction at the other $\pi NN$ vertex, drives the 
value of the neutron's electric dipole moment. As standard in
such assays, we assume the magnitude of the 
external momentum transfer $|q|$, quark masses, and meson masses
all small compared to the nucleon mass $M$. 
We wish to connect this language to the 
Fock-state expansion of the light-front 
formalism, realized in terms of the fundamental quark and gluon 
degrees of freedom. 
The chiral Lagrangian framework
allows us to estimate the contribution of the $udd u {\bar u}$
Fock component of the neutron to its electric dipole
and anomalous magnetic moment. Comparing the CP-conserving
and CP-violating $\pi-N$ loop graphs of Fig.~\ref{fig:edmloop}, 
we estimate 
\begin{equation}
|\beta_a| \approx 2 \frac{|\bar g_{\pi NN}|}{|g_{\pi NN}|} 
\log(M_N/M_\pi) \approx
4 \left(\frac{0.027}{13.4}\right) |\bar \theta| \,,
\end{equation}
where $\bar g_{\pi NN}$ is the CP-violating $g_{\pi NN}$ coupling
constant. We use the numerical estimate of Ref.~\cite{Crewther:1979pi}
for $\bar g_{\pi NN}/g_{\pi NN}$ and include the well-known
logarithmic enhancement of the CP-violating graphs in $\beta$,
as such is absent in the CP-conserving analog~\cite{Jenkins:1992pi,MeisStein}. 
Employing Eq.~(\ref{unirel}) and 
assuming the $udd u {\bar u}$ Fock component dominates both the
anomalous magnetic and electric dipole moment, 
we find
\begin{equation}
d^n \sim \bar \theta 
e (2\cdot 10^{-16} \hbox{cm}) \,,
\end{equation}
which is roughly a factor of 2 smaller than the 
estimate of Crewther et al.~\cite{Crewther:1979pi} but just that
computed by Ref.~\cite{Pospelov:1999ha} employing the QCD sum rule 
approach. 
The value of $d^n$ can also be computed 
within the framework of 
lattice 
QCD~\cite{Aoki:1989rx,Guadagnoli:2002nm,Shintani:2005xg,Berruto:2005hg}. 
Both estimates are compatible with a recent computation of $d^n$
employing dynamical light quarks~\cite{Berruto:2005hg}. 
It is worth noting that the assumptions underlying the two estimates
are slightly different. 
Our estimate also follows if 
the fractional contributions of the $\pi-N$ loop
graph to the electric dipole moment and to the anomalous magnetic
moment are the same; they need not be numerically dominant. 
Indeed, detailed analyses of the magnetic moments show that 
the $\pi-N$ loop graphs are not numerically 
dominant~\cite{MeisStein,Puglia:1999th}. 
In comparison, the $d^n$ estimate of Crewther et al.
follows from computing the ``long-distance'' $\pi-N$ loop graph
and assuming it numerically dominant. 
In the chiral, $M_\pi \to 0$, limit, this
is seemly, as the 
$d^n$ estimate contains an explicit factor of $\log(M/M_\pi)$. 
The numerical dominance of this contribution is less
clear for physical values of the $\pi$ mass, though our own
estimate, stemming from a different assumption, is compatible with
that of 
Crewther et al. With our assumptions we also conclude that 
$d^p \sim - d^n$, so that we predict that the 
electric dipole moment of the nucleon is predominately isovector. 
Indeed, the isospin structure of the $\pi-N$ loop diagrams precludes
any isoscalar contribution; we assume these contributions drive
that of the $udd u \bar u$ Fock state.  
Turning to the $q^2$ dependence of the electric dipole and 
anomalous magnetic form factors, the structure of our Eq.~(\ref{genrel})
shows that the $q^2$ dependence of $F_3(q^2)$ ought track
that of $F_2(q^2)$, Fock state by Fock state, as we have argued
on general grounds that $\beta$ should be independent of $q^2$. 
Moreover, if we model the contribution of the 
$ddu u\bar u$ Fock state through the $\pi-N$ loop graph as per
Fig.~\ref{fig:edmloop}, as we have discussed in the $q^2=0$ limit,
we observe, under our stated assumptions, that
the isospin structure of $F_3(q^2)$ 
is also isovector. This, too, follows from
the isospin structure of the $\pi-N$ loop graphs. Our analysis
is at odds with one conclusion of Ref.~\cite{Hockings:2005cn}, 
as its authors find the $q^2$ dependence of $F_3(q^2)$ unlike that
of the other electromagnetic form 
factors~\cite{Bernard:1992qa,Bernard:1998gv}. 
As in the $q^2=0$ limit, 
our prediction does not require the $\pi-N$
loop graphs to dominate the $F_3(q^2)$ form factor.

We can also estimate $\beta$ through 
CP violation in the weak interaction, as mediated through
the CKM matrix. The value of $d^n$ through this mechanism of
CP violation is much smaller, as 
it first appears in 
${\cal O}(G_F^2 \alpha_s)$~\cite{Shabalin:1978rs,Shabalin:1980tf}. 
Here, following Ref.~\cite{Khriplovich:1981ca}, we anticipate that the dominant
contribution to the electric dipole moment 
of the nucleon is mediated by a hadronic
loop graph with a $\pi \Sigma$ intermediate state, as illustrated
in Fig.~\ref{fig:edmloop}c) for the neutron. Note that two diagrams
contribute, each with a single CP-violating $N\pi\Sigma$ vertex. 
Since the $ddu s\bar u$ Fock state makes a negligibly small contribution
to the neutron's magnetic moment, we proceed to estimate $d^n$ by summing
Eq.~(\ref{genrel}) over all Fock states for $q^2=0$ 
and assuming that diagrams
akin to Fig.~\ref{fig:edmloop}c)  do drive its numerical value. 
Using Eq.~(\ref{start}) and
writing 
\begin{equation}
\beta \approx 2 G_F^2 \alpha_s(1\,\hbox{GeV}) J_{\rm CP} M_\pi^3  \,,
\end{equation}
noting the Jarlskog invariant~\cite{Jarlskog:1988ii},
$J_{\rm CP} \sim 3\cdot 10^{-5}$, and 
$\alpha_s(1\,\hbox{GeV})\approx 0.3$, 
we find
\begin{equation}
d^n \sim e 3.6 \cdot 10^{-32} \hbox{ cm} \,,
\end{equation}
which is rather comparable to the estimate of Ref.~\cite{Khriplovich:1981ca},
namely,
\begin{equation}
d^n \sim e 2 \cdot 10^{-32} \hbox{ cm} \,,
\end{equation}
which follows from a direct estimate of the chirally enhanced terms. 
This procedure also implies that $d^p \sim - d^n$, yielding an
isovector electric dipole moment. 
The electric dipole moments have also been computed in chiral
perturbation theory in the context of the factorization hypothesis, 
yielding~\cite{Bijnens:1996ni}
\begin{equation}
d^n \approx \pm 5.3 \times 10^{-32} \hbox{e-cm} \quad;\quad
d^p \approx \mp 3.6 \times 10^{-32} \hbox{e-cm} \,, 
\end{equation}
where the manifest sign of $d^n$ and $d^p$ is not determined. 
These authors also discuss the connection to the anomalous
magnetic moments in the context of their approximations. 
Note, too, that the dominantly isovector nature of the electric dipole
moments is manifest in their results.

\section{Conclusions}

We have derived exact formulae for the electromagnetic form factors, 
$F_1(q^2)$, $F_2(q^2)$, and $F_3(q^2)$ of the nucleon, and indeed
for all spin-$1/2$ systems,  
in the light-front formulation of quantum field theory, thus extending
the treatment of Ref.~\cite{BrodskyDrell}
to the analysis of the time-reversal and parity-odd observable
$F_3(q^2)$. To realize this we have developed the light-front representation of discrete
symmetry transformations, ${\cal T}_\perp$, ${\cal P}_\perp$, and
${\cal C}$ and have shown how ${\cal T}_\perp$-odd and ${\cal P}_\perp$-odd 
effects 
can be represented by the phases of light-front wave functions.
The explicit expressions
which we have developed for $F_2(q^2)$ and $F_3(q^2)$ have the desired 
transformation properties under ${\cal T}_\perp$, ${\cal P}_\perp$, and
${\cal C}$.  
As a result, 
we find a 
universal relation between $F_3(q^2)$ and $F_2(q^2)$, Eq.~(\ref{genrel}), 
Fock state by Fock state, which follows independently of the 
mechanism of CP violation at the Lagrangian level.   

We have employed our
relation to estimate the electric dipole moments of the nucleon
through both strong and weak interaction CP violation in the
Standard Model and find results comparable to existing
estimates. We find that the relation $d^n \sim - d^p$ 
emerges on rather general grounds, echoing the  
isospin structure 
of the empirical anomalous magnetic moments, $\kappa^n \sim - \kappa^p$.

\noindent{\bf Acknowledgment}\,\,
S.G. thanks the SLAC theory group for hospitality during the
completion of this project. 
%is supported by
%the U.S. Department of Energy under contracts 
%DE-FG02-96ER40989 and DE-FG01-00ER45832.  

\section{Appendix: Conventions}

We employ the Dirac representation for $\gamma^\mu$: 
\begin{equation}
\gamma^0=\left(
\begin{array}{cc}
I &0\\
0&-I
\end{array}
\right)
\ ,\
\gamma^i=\left(
\begin{array}{cc}
0&{\sigma}^i\\
-{\sigma}^i&0
\end{array}
\right)
\ ,\
\gamma_5=\left(
\begin{array}{cc}
0&I\\
I&0
\end{array}
\right) \ ,
\label{a1}
\end{equation}
where ${\sigma}^i$ are Pauli matrices, $I$ is the $2\times 2$ unit matrix, 
${\sigma}^{\mu\nu}={i\over 2}[{\gamma}^{\mu},{\gamma}^{\nu}]$, 
$\gamma_5=i \gamma^0 \gamma^1 \gamma^2 \gamma^3$, and
$\gamma^\pm \equiv \gamma^0 \pm \gamma^3$. 
For the light-cone spinors $u(p,\lambda)$ and 
$v(p,\lambda)$, we use~\cite{Lepage:1980fj}
\begin{equation}
u(p,
+1)={1\over {\sqrt{2p^+}}}
\left(
\begin{array}{c}
p^++m\\
p^R\\
p^+-m\\
p^R
\end{array}
\right)
\ ,
\quad
u(p,
-1)={1\over {\sqrt{2p^+}}}
\left(
\begin{array}{c}
-p^L\\
p^++m\\
p^L\\
-p^++m
\end{array}
\right)
\ 
\label{ss2}
\end{equation}
and
\begin{equation}
v(p,
+1)={1\over {\sqrt{2p^+}}}
\left(
\begin{array}{c}
-p^L\\
p^+ - m \\
p^L \\
-p^+ - m 
\end{array}
\right)
\ ,
\quad
v(p,
-1)={1\over {\sqrt{2p^+}}}
\left(
\begin{array}{c}
p^+ - m \\
p^R\\
p^+ +  m \\
p^R \\
\end{array}
\right)
\ ,
\label{vss2}
\end{equation}
where
we define 
$p^R\equiv p^1+ip^2$,
$p^L\equiv p^1-ip^2$, and 
$p^{\pm}\equiv p^0\pm p^3$.
Moreover, we employ the notation
$k^\mu = (k^+, k^-, k^L, k^R)$ so that 
$k \cdot x = (1/2)(k^+ x^- + k^- x^+ - k^L x^R - k^R x^L)
= (1/2)(k^+ x^- + k^- x^+) - \mathbf{k}_\perp\cdot \mathbf{x}_\perp$.

\end{document}